\pdfoutput=1
\documentclass{sigchi}

\usepackage{balance}  
\usepackage{graphics} 
\usepackage{txfonts}
\usepackage{times}    
\usepackage[pdftex]{hyperref}
\usepackage{url}      
\usepackage{color}
\usepackage{textcomp}
\usepackage{booktabs}
\usepackage{ccicons}
\usepackage{todonotes}

\makeatletter
\def\url@leostyle{%
  \@ifundefined{selectfont}{\def\UrlFont{\sf}}{\def\UrlFont{\small\bf\ttfamily}}}
\makeatother
\def\pprw{8.5in}
\def\pprh{11in}

\definecolor{linkColor}{RGB}{6,125,233}
\hypersetup{%
  pdftitle={SIGCHI Conference Proceedings Format},
  pdfauthor={LaTeX},
  pdfkeywords={SIGCHI, proceedings, archival format},
  bookmarksnumbered,
  pdfstartview={FitH},
  colorlinks,
  citecolor=black,
  filecolor=black,
  linkcolor=black,
  urlcolor=linkColor,
  breaklinks=true,
}

\begin{document}

\title{Exploring the Front Touch Interface for Virtual Reality Headsets}


\numberofauthors{2}
\author{
  Jihyun Lee$^{1,2}$ \and Byungmoon Kim$^1$ \and Bongwon Suh$^2$ \and Eunyee Koh$^1$ \\
  \alignauthor{
  $^1$\affaddr{Adobe Systems, Inc.} \\
    \affaddr{San Jose, CA} \\
    \email{\{bmkim, eunyee\}@adobe.com}
  }
  \alignauthor{
  $^2$\affaddr{Seoul National University} \\
    \affaddr{Seoul, Rep. of Korea} \\
    \email{\{jihyunlee, bongwon\}@snu.ac.kr}
  }
}

\maketitle

\begin{abstract}
In this paper, we propose a new interface for virtual reality headset:
a touchpad in front of the headset.
To demonstrate 
the feasibility of the front touch interface, we built a prototype device, explored VR UI design space expansion, 
and performed various user studies.
We started with preliminary tests to see how intuitively and accurately people can interact 
with the front touchpad. Then, we further experimented various 
user interfaces such as a binary selection, a typical menu layout, and a keyboard.
Two-Finger and Drag-n-Tap were also explored 
to find the appropriate selection technique. 
As a low-cost, light-weight, and in low power budget technology, a touch sensor can make an ideal interface for mobile headset. Also, front touch area can be large enough to allow wide range of interaction types such as multi-finger interactions. 
With this novel front touch interface, we paved a way to new virtual reality interaction methods. 
\end{abstract}

\keywords{Virtual Reality; Input-interaction; Proprioception.}

\category{H.5.2.}{Information Interfaces and Presentation (e.g. HCI)}{User Interfaces - Input devices and strategies}

\section{introduction}
Virtual Reality (VR) refers to three-dimensional realities implemented with
stereo viewing goggles and reality gloves \cite{AR}. The immersive
experience that the VR technology provides entails a wide range of potential
usages, attracting attentions from both the industrial and research fields for
the last decades, but still has a room for improvement \cite{VRhuge}. 
``Since its inception, the field of VR has revolved around the head-mounted displays (HMD)"
\cite{FutureVR}. For the last several years, industry efforts to make VR headsets 
more accessible to the public could be noticed by the development of mobile VR headsets such as Samsung/Oculus's Gear VR,
Zeiss's VR One and Google's Cardboard. Mobile VR headset has shown its potential
to be the next big platform, and become as ubiquitous as their power system -
smartphone - with its low cost, light weight and high portability
characteristics. VR has never been as accessible and afforable as now.

\begin{figure}[!t]
\centering
\includegraphics[width=0.75\columnwidth]{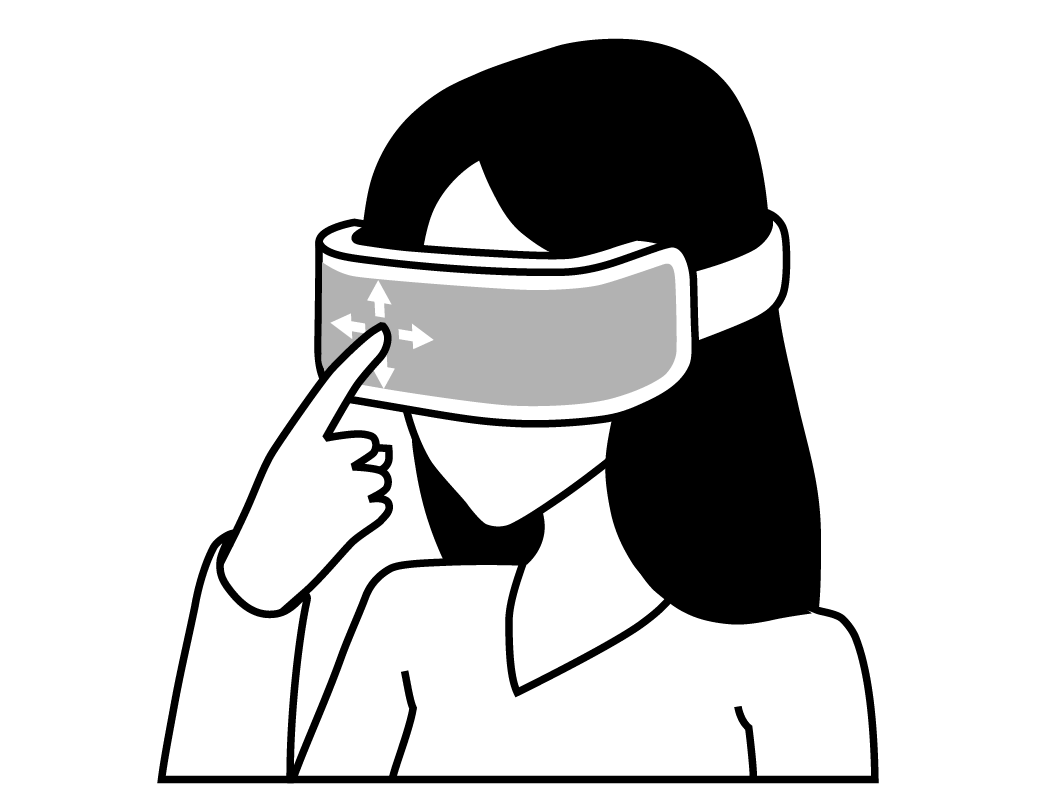}
\caption{A concept art of the front touch interface for virtual reality headset.
Users use their proprioceptions to locate the VR objects with respect to their 
bodies by touching on the touchpad placed around the face. }
~\label{fig:concept-art}\vspace{-6ex}
\end{figure}

Upsurging use cases of VR as a medium for movie-watching \cite{MovieSilver}, photo-curating and reading 
\cite{NarrVR} besides its traditional game application or medical training denote its 
potentials to infiltrate our daily lives. More and more mass media applications 
such as Netflix (Figure \ref{fig:netflix}), Hulu, and Twitch are continuously 
being introduced to VR headsets \cite{massmedia}. Facebook's acquisition of Oculus
highlighted the excitation of the media that VR device might be the next big
platform of social network \cite{FacebookArticle}. 
With VR being never as accessible
and affordable as now, it is an ideal time to evolve VR as the next
communication interface and establish a new metamedium \cite{CommVR}.
For VR headset to support mass media application for a wide range of 
people, its interaction method should be easier for generic tasks 
such as menu selection and keyboard application (Figure \ref{fig:netflix}). 

We, therefore, propose a front touch interface for an intuitive, easy, and practical VR interaction. 
Here, people use the touchpad placed in front of the headsets (Figure \ref{fig:concept-art}) to interact with 
virtual environment. Their proprioceptions help them to intuitively locate virtual items they see in front.
The physical contact with the pad also helps users manipulate the UI events 
more easily and precisely \cite{MOIS}. The low-cost, low-weight and low power budget (Table \ref{tab:power})
characteristics of touch sensing technology makes this interface a practically 
viable and portable solution, especially for mobile VR headsets.

We implemented a prototype device to explore various design options and perform user studies. 
The front-touch interface enables the VR cursor be weakly tied to the gaze, and the windows to 
be fixed to the head movement. We conducted preliminary studies on intuition and accuracy of the 
proprioception, and a short study to compare four different design options.
We then performed two formal user studies on menu selection and text-input using VR keyboard.
We summarize major contributions in this paper as: 
\begin{itemize}
\item Proposed the front-touch interface for VR headset and built a prototype device.
\item Expanded the VR interaction design space.
\item Introduced new front-touch interaction methods: Two-Fingers and Drag-n-Tap.
\item Evaluated the front touch interface by conducting user studies on intuition, accuracy, design space validation, menu selection, and text-input.
\end{itemize}
\begin{table}[!t]
  \centering
  \begin{tabular}{r c c}
    \toprule
        & {\small \textbf{Active Mode (W)}} & {\small \textbf{Inactive Mode (W)}} \\
    \midrule
    	{\small \textbf{Cellphone Touch Sensor}} & 21.0m $\sim$ 30.0m & 60.0$\mu$ $\sim$ 90.0$\mu$ \\
	{\small \textbf{Leap Motion \cite{dim}}} & $3\sim5$ & --- \\
    \bottomrule
  \end{tabular}
  \caption{Power requirement of touch sensor could be very low compared to motion capture using 
camera (Leap Motion). }~\label{tab:power}\vspace{-4ex}
\end{table}

\section{related work}
There are a plethora of research and industry products on virtual reality devices and interactions. 
However, an ability to provide precise and intuitive input signal is still identified as an issue among the VR users \cite{DOS}.
In this section, we address how prior studies approached to solve selection tasks, especially with the menu interface, and text-entry.

\subsection{Interaction Methods and Devices}
Bowman \cite{FormalizeVE}
characterized the four universal interaction tasks of virtual environment systems: selection, navigation, manipulation, and system control.
Although simple selection would be adequate for static menu layouts, 
delicate manipulation is needed for complex UI where buttons are smaller and packed.

\subsubsection{Gestural Interaction}
The most obvious selection technique in VR would be the gestural interaction
where users use their body to interact with VR objects \cite{constraints, grabandmani}.
It provides a natural user interface in VR as they select objects using their proprioceptive senses \cite{MOIS}. 
Ray-casting \cite{vmenu1, constraints4r, grabandmani} even lets users to select distant items by 
using a ray extended from user's hand allows users to interact with distant items.
Often it is accompanied with other VR devices such as gloves \cite{vmenu3} to allow more delicate inputs and enhance the user experience. 
Although motion capture technology using camera or photosensors (mounted on the headset \cite{DOS21r}
or in a room-scale) has advanced, it still suffers from technical issue to serve high precision \cite{toms}.

\begin{figure}[!t]
\centering
  \includegraphics[width=\columnwidth]{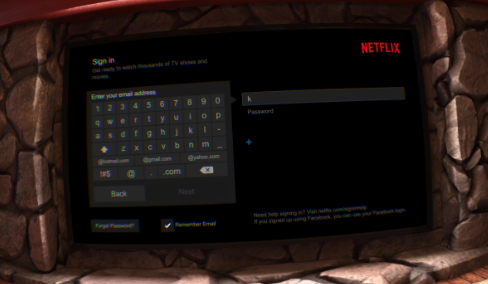}
  \caption{A recently released (i.e. 2015-09-24) VR application for media consumption 
  with a keyboard interface. Such mobile VR application will be used ubiquitously often without 
  additional peripheral devices nearby.}~\label{fig:netflix}
  \vspace{-3ex}
\end{figure}

\subsubsection{Gaze-based Interaction}
Extensive amount of research has been done on the usability of gaze-based 
interaction in head-mounted displays \cite{eyetrackhci3, eyetrackhci, eyetrackhci2}. It tracks the eyes or the head orientation \cite{VEIT} to calculate the pointing direction. A crosshair positioned at the finger tip or at the center of the gaze \cite{gazenotbetter, VEIT} helps to indicate where the selection will be made. Selection command is made by dwelling the gaze on item \cite{gazenotbetter} or other standard input signals such as pressing a physical button\cite{VEIT, MOIS}. 
Mobile VR headsets in the market tend to prefer the head-oriented selection technique over the traditional input methods since 
it provides more simple and intuitive interaction \cite{VEIT, MOIS}. 
Selection commands are made by tapping on the touchpad (Gear VR) or by pulling a magnet button (Google Cardboard) positioned at the side of the headset. 

\subsubsection{Handheld or other Assisting Devices}
Other interaction types include but not limited to gloves \cite{vmenu3}, pen and tablet \cite{vmenu3r12r, handheld2dto3d}, 3D mouse, standard keyboard, and other handheld indirect props \cite{korean}. 
3D mouse especially is appropriate for more delicate VR works such as industrial product design due to its resemblance of the computer mouse \cite{vmenu3r8r, korean,vmenu3r7r}. In \cite{vmenu3}, interaction with menus using the gloves was compared with pen and tablet menus, and floating menus. Its performance was less impressive but opened a new design space for VR menu selection. We also have speech recognition \cite{vmenu2, constraints16r, vmenu1}. But both handheld devices and speech recognition have rooms for improvement to be 
the mainstream of the VR input methods \cite{constraints}.

\subsection{Design for Virtual Menus}
Just like the 2D space menus, various designs of virtual menus have 
been experimented. There has been an attempt to apply the
desktop metaphor on VR, but it is often asserted that VR needs its own metaphor for optimal design \cite{vmenu3r6r}. In 
\cite{wimp}, for instance, proposed a design option of using widgets instead of WIMP in virtual environment.

\subsubsection{Relative Positions}
Menu items can be floating in the air \cite{vmenu1}, 
or positioned relative to the user's head \cite{vmenu3r8r, orbitalmode}, hand \cite{vmenu3} or whole body \cite{MOIS}.
\cite{handheld2dto3d} coins them as world-fixed windows for menus at their absolute position in VR, 
and view-fixed window for menus that follow the head movement 
(\textit{surround-fixed} and \textit{display-fixed} in \cite{handheld2dto3d7r}).
Pros of having the menus relative to the body is that 
users can take advantage of their proprioceptive cues during the interaction \cite{vmenu3}, 
and have menus always within sight. 
It is asserted that exploiting proprioceptive feedback helps users to perform better 
than just relying on visual feedback \cite{MOIS}. 
This design option, however, cannot be used with head-oriented interaction 
which has a cursor fixed at the center of the window. 

\subsubsection{Layouts}
The layout of menu items largely depends on the purpose of using VR. 
For instance, the realistic quality would be the focus for training while 
the usability would be the priority for mass media applications where 
a series of rudimentary tasks would be conducted \cite{vmenu3}.
Various layout designs have been proposed to support high usability, 
which include drop-down \cite{constraints7r, vmenu2, vmenu1} and circular layouts \cite{holosketch}.
In general, menu system takes advantage of its one degree-of-freedom operation \cite{vmenu3, vmenu3r7r, vmenu3r15r} to
enable simple interaction.
3d widgets have also been tried out to give affordance but suffer from lower precision as the 
menu items have volumetric shape \cite{VEIT} 

\subsection{Text Entry}
Text entry 
is an integral component of any advanced applications (Figure \ref{fig:netflix}). 
VR keyboard application has been tested with various means such as gestural interactions with gloves \cite{bowman2002novel}, 
tablet + stylus combination, \cite{DOS7r} and a standard keyboard in mixed reality \cite{DOS}. 
Gaze-controlled interaction was also found to be viable for the motor- or speech- impaired users \cite{ gaze4impaired2, gaze4impaired}.
However, gestural interactions suffer from high power budget and cannot support delicate inputs yet. Also, interactions using external devices supports low portability, and restricts the distinguishing characteristic of the \textit{mobile} VR headsets.

Within our best knowledge, no interface with a touchpad in front of VR headsets has been proposed, let alone evaluated.
We first started with a short preliminary study to test how intuitive the front touch interaction is.

 
\section{Pre-Study1: Intuition}
While using VR headsets, users sometimes click the front of the
headset (click what you see) even though they know that the touchpad is on the
side of the headset not on the front. So, we conducted an informal user study to
inquire users about what interaction techniques are most
intuitive to click one of two rectangle objects in the VR scene. For this short pilot user
study, we recruited 10 study participants who have never used VR headset before
and who do not have any prior knowledge about latest VR input technologies, side-touch on VR headsets, or Google cardboard. Study participants include 9 males and 1 female, and the age ranges are four participants in 20-30, four
participants in 30-40, and two participants in 40-50.

\begin{figure}[!t]
\centering
  \includegraphics[width=\columnwidth]{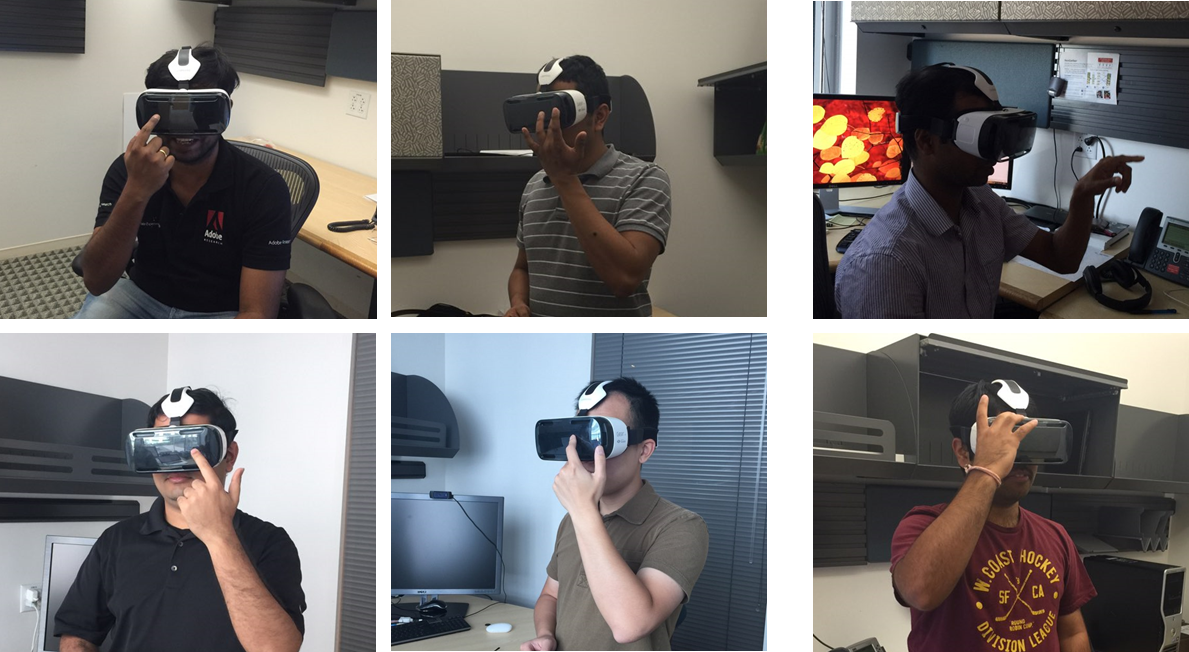}
  \caption{
  All participants had no prior VR experience. To interact with VR objects as their first attempts, 
  in the first two columns, people touched on the front of the headsets. The top right shows a case of mid-air, and the bottom right shows 
  a case of touching the edge of the headset. Seven out of ten participants responded with the front touch to interact. }~\label{fig:intuition}\vspace{-3ex}
\end{figure}

We helped the participants wear the headset so that they cannot see the side touch on VR.
In the VR scene, participants could see two rectangle objects, one in red and the other in blue. 
After we checked with participants whether they can see those two rectangle objects, 
we asked a following open question: ``how would you interact with the
red rectangle that you see in the VR scene?". From 10 participants, 
five participants said front-touch on the headset, three said mid-air gesture, and 
two said side-touch on the headset. 
Figure \ref{fig:intuition} shows participants' gesture while answering this question. 
For three study participants who mentioned mid-air
gesture, we asked second question: ``what if the input technology is on
the headset, where can be the most intuitive place?". Among the three,
two participants said front of the device and one participant said side of the
device. From this study, 7 participants said it is intuitive to have the front
input device while 3 participants said somewhere on the side of VR headset. This
pilot study results show that people had a natural inclination to use 
their proprioceptions to touch what they see in front by placing their hands onto 
the front of the headsets.
And the front of the headset is
more intuitive place for study participants than the side of the headset for the input device on the VR headset.
To conduct more advanced experiments, we constructed a prototype device 
for the front touch interaction.

\section{implementation}

\subsection{Prototype Device}
We used Samsung/Oculus Gear VR for S6 and Samsung Galaxy S6
smartphone for the VR headset, and Samsung Galaxy Note4 as the touch sensor. 
Gear VR uses gaze-directed selection technique and a side touchpad to 
signal the selection \cite{samsung3, samsung1} as its built-in interface.
As a prototype device, we physically tied the Note4 behind the S6. 
The touch events from Note4 were then transmitted to S6 via bluetooth. 
Note4 supports a resolution of 2560x1440-pixels, and thus the $(x, y)$ point on the 
touch sensor ranged from ([0,2559], [0,1439]).
The touch sensor will send what kind of touch action (e.g. touch-down, scrolling, lift-up) 
was done and the touch point coordinations $(x, y)$ to the VR system, so that 
corresponding changes can occur in the scene. 
It was only a prototype, and we envision the ultimate interface to resemble Figure 
\ref{fig:concept-art}, where the touch sensor is embedded to the headset, covering a wider range of face.

\begin{figure}
\centering
  \includegraphics[width=\columnwidth]{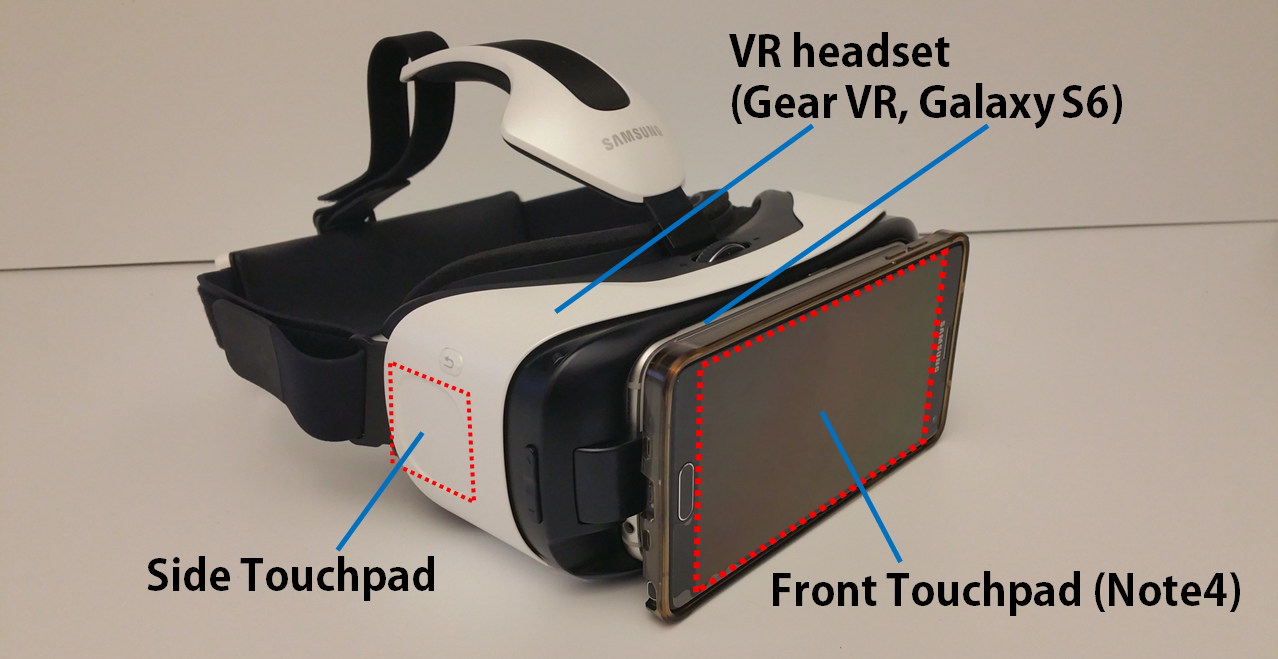}
  \caption{An apparatus setup for a prototype device used throughout the experiments.
  We physically tied a smartphone (Note4) at the front of the headset, and used 
  it as a touch sensor. Since we used a smartphone as a touch sensor, we had 
  an extra weight penalty and occasional bluetooth connection instability.}~\label{fig:device}\vspace{-4ex}
\end{figure}

\subsection{VR Scene Graph and Picking For UI Events}
We implemented a simple scene graph engine using OpenGL ES 3.0, Android NDK (Native Development Kit), 
and Oculus SDK. We implemented standard material shading model (Blinn-Phong) for rendering. 
Buttons and cursors and other shapes were generated procedurally. In addition,
we used the STB library to generate text geometry. To generate UI events, we
need to know which objects are intersecting with the ray starting from the scene
camera and pointing the cursor. We also should know which intersecting object is the closest. This is the object picking
problem. Since VR renders two scenes for two eye locations that are generated
from the scene camera, by displacing the camera location to the left and right,
the scene camera location is equivalent to the center between the two eyes. The
rest of picking is straight forward. We march from the root of scene graph
clipping the ray by bounding box of each node. If we hierarchically organize UI elements into the scene graph, 
hierarchical bounding boxes are generated automatically. Since all nodes that belong to
a bounding box will not be visited when the ray does not intersect with the bounding box, we can save lots of computations.
This way, we can accommodate a large number of UI elements in the scene.
If the ray clipped by bounding box is not empty, we perform full ray/triangle intersection test. If the
object, i.e., UI button, intersects with the ray, we report the button object so that
a UI event associated with that button can be generated.

\subsection{Matching Coordinations between Touch sensor and VR}
To construct equations that map the 2D coordination system of the touchpad 
to 3D VR, we collected the data from the actual users.
Ten participants were given a cross-shaped object positioned randomly 
on a 13-by-9 grid in VR scene, and were asked to touch on the front pad 
to ``select" on it.  The grid was designed by varying the 
horizontal angle ($\theta_1$) and the elevation angle ($\theta_2$) from the viewpoint.
Each participant made $13 \times 9 \times (6 \text{ sessions}) = 702$ selections. 
With this data, we performed linear regression to model the equations. 
The linear correlation coefficient for $x$ and $\theta_1$ was 0.998; 
that for $y$ and $\theta_2$ was 0.989. 
We also measured the central point (the sample mean of the observed points) for 
each cross, and found the average distance between the central point and each observed point 
to be $\approx 184$ pixels. 

From these brief analyses, we could induce that people are reasonably good at interacting with 
the front touch using their proprioceptions without visual cues of their hands.
However, since there is an offset of 184 pixels when trying to select the same target, 
well-designed selection control mechanism was needed. 


\begin{figure}
\centering
\includegraphics[width=\columnwidth]{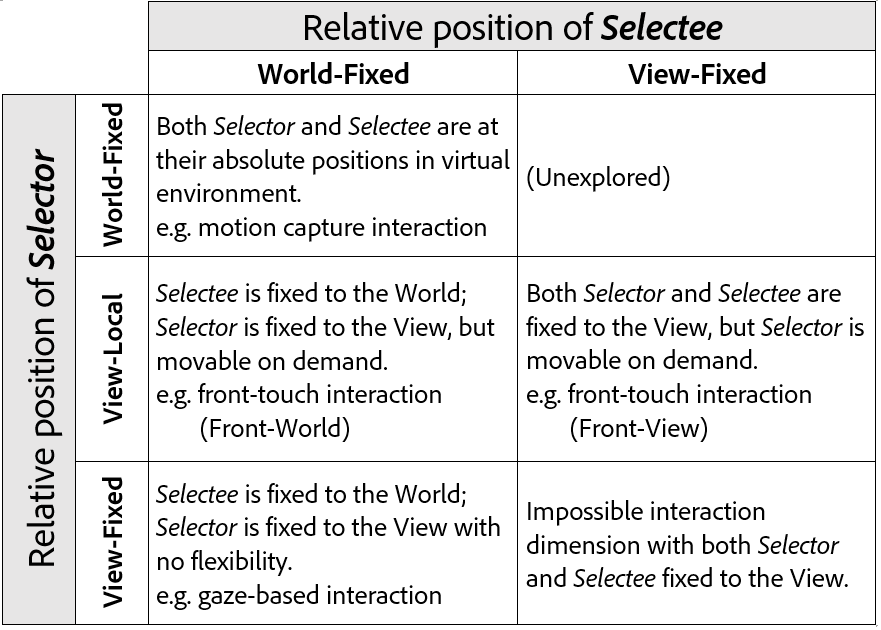}
\caption{Design space matrix based on the relative placements of \textit{selector} and \textit{selectee}. 
Note that, in gaze-based interaction, since \textit{selecter} is fixed at the center of the scene,
\textit{selectees} must stay fixed in world space to allow users to look at \textit{selectees}. 
However, since users can move \textit{selector} in front touch interaction (i.e. view-local),
we can now make \textit{selectees} to move with the view (head). This allowed the design space expansion. 
}~\label{tab:design-space}\vspace{-4ex}
\end{figure}

\section{design space}
Expanding on the concepts of \textbf{world-fixed} and \textbf{view-fixed} window \cite{handheld2dto3d}, 
we constructed a design space that layouts the relationships between the \textit{selector} and the \textit{selectee} (Table \ref{tab:design-space}).
\textit{Selector} is the selecting subject, such as users' fingers in gestural interface, the tip of stylus on tablet or
a virtual cursor. \textit{Selectee} is the selecting object such as the VR objects that users intend to interact with.

View-fixed window cannot be supported for the gaze-based interaction since the 
cursor is always fixed at the center of the sight.
By introducing the front touchpad, we could naturally expand the design space, 
and make the cursor movable relative to the View.
Users can liberate the cursor from the head movement if they want to use their fingers 
and work in an eye-off manner \cite{vmenu3, MOIS}. 
The middle row in Table \ref{tab:design-space} demonstrates this expansion of design space 
with the introduction of the front touch. 
Also, since this interface can be embedded to the headset with low cost, 
it makes it suitable for mobile VR headsets.

\section{Pre-Study 2: Explore the Design Space}
We explored three design spaces 
with four different interaction methods using the built-in side touchpad of the headset, 
and the newly proposed front touch interface. The bold text is the name of an interaction method, 
and text in a bracket is its corresponding design space as (\textit{selector}, \textit{selectee}).

\textbf{Side-Gaze} (View-Fixed, World-Fixed) is a gaze-based interaction where people use their heads to 
move view-fixed cursor onto world-fixed VR objects. They tap on the side touchpad to signal the selection.\\
\textbf{Front-Gaze} (View-Fixed, World-Fixed) is the same as Side-Gaze except that users tap on the front touchpad instead of the side touchpad.\\
\textbf{Front-World} (View-Local, World-Fixed) is a selection method we tested to find the feasibility of the front 
touch interaction. VR cursor is fixed to the view, but we can move its position in the scene by using the front touchpad.\\
\textbf{Front-View} (View-Local, View-Fixed) is another front touch method with \textit{selectee} relatively positioned to the view. 

Experiment was conducted with a user interface that resembles a typical binary UI with two options (e.g. yes / no).
Here, two equally-sized planes (left and right) divided by a center vertical line was given, where one plane was red and the other blue. 
Each task begins with a participant selecting the red. Once s/he selects the red plane to begin the task, the selected plane becomes blue and the other becomes red for selection. Each experiment was composed of a series of 20 tasks, and 18 participants (13 males, ages ranging from 20's to 40's, all right-handed) were recruited.
We measured how quickly and accurately participants can select from one plane to another.
In total, we could collect $\text{20 tasks} \times \text{4 methods} \times \text{18 participants}
= \text{1440 selections}$. Both the choice of the starting the red plane and the order of selection methods were randomized.

\begin{figure}[!t]
\centering
  \includegraphics[width=\columnwidth]{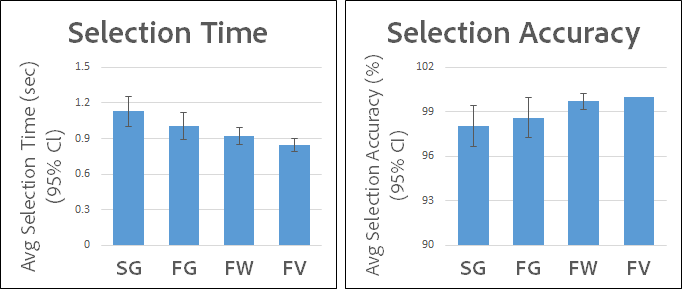}
  \caption{Average selection time and accracy of each method: Side-Gaze (SG), Front-Gaze (FG), 
  Front-World (FW) and Front-View (FV). Front-View where the front touch interaction with View-fixed 
  scenes was used outperformed in both selection time and accuracy.}~\label{fig:2-time-acc}
\end{figure}

\subsection{Performance Results}
As shown in Figure \ref{fig:2-time-acc}, participants were quickest and most accurate in the order of Front-View, Front-World, Front-Gaze and Side-Gaze. We performed one-way repeated-measures ANOVA on both the average selection time ($F_{3,51} = 9.658, p < 0.00005$) and accuracy ($F_{3,51} = 260.764, p < 0.05$) and found significant differences among the tested interaction methods. 
Performance results reveal that users can make selections more quickly and accurately with view-local \textit{selector} in a relatively simple user interface (binary selection test). It was especially high performing when the UI elements were fixed to the view as well (Figure \ref{fig:2-time-acc}). To understand better on the front touch with more complex user interfaces, we conducted in-depth experiments to compare Side-Gaze and two other refined front touch techniques.


\section{study 1: Selection Test}
In this study, we tested three different selection techniques under a
typical menu interface in VR (see Figure \ref{fig:2-sshot}): Side-Gaze (i.e. Side), Two-Fingers, and Drag-n-Tap.
Side is the same technique tested in the previous study, which used the side touchpad and gazing.
Two-Fingers and Drag-n-Tap are techniques that we designed for advanced front interaction 
to overcome precision issue due to the lack of visual cue information on the finger position until a touch is made.

\subsection{New Selection Techniques for Front Touch}

\subsubsection{Two-Fingers}
In this technique, a pair of fingers work together on the front touchpad:
one finger drags the VR cursor by moving across the front 
touchpad, and with this dragging finger still on the touchpad, 
the other finger makes a tap to initiate the selection. 

\subsubsection{Drag-n-Tap}
In this technique, one finger does both the moving and tapping.  Users will 
move the cursor by dragging it on the front touchpad, and 
lift their fingers from the pad to tap to signal a selection.

\subsection{Study Design}
\begin{figure}[!t]
\centering
  \includegraphics[width=\columnwidth]{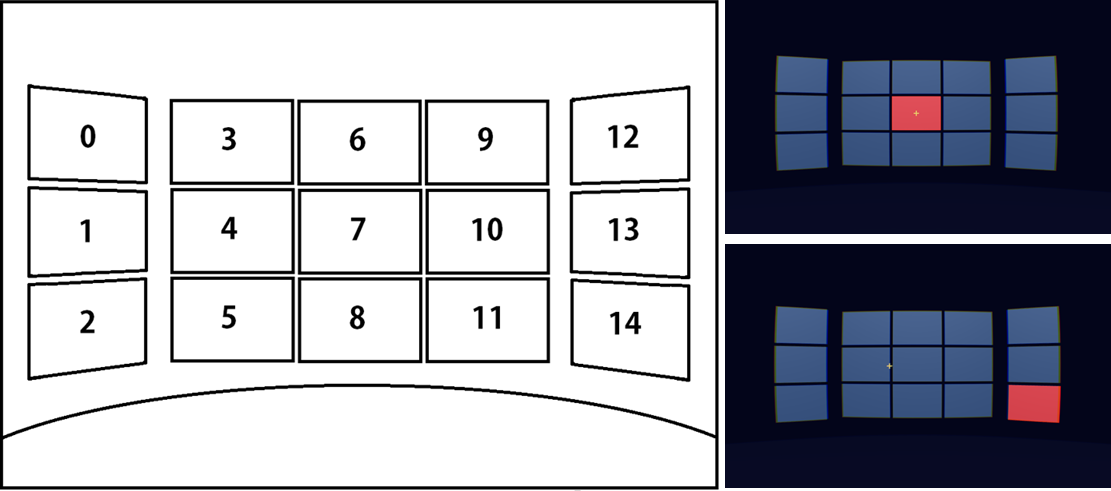}
  \caption{Left image illustrates the user interface used for the experiment, and the number assigned to each button. 
  Right two images are sample snapshots; the top image shows a case where the button no.7 is the target, and the
  button a case where the button no.14 is the target. }~\label{fig:2-sshot}
\end{figure}

\begin{table}[!t]
  \centering
  \begin{tabular}{p{.95\columnwidth}}
  \toprule
  \centering \small\textbf{Post-Questionnaires}\tabularnewline
    \midrule
\small{1. How easy was it to learn the technique?} \\
\small{2. How easy was it to locate the touchpad?} \\
\small{3. How efficient was this technique at completing the given task?} \\
\small{4. How confident were you that the correct target button will be selected?} \\
\small{5. How tired did your \underbar{arm} become while using this technique?} \\
\small{6. How tired did your \underbar{neck} become while using this technique?} \\
\small{7. Did you experience nausea/dizziness while using this technique?} \\
\small{8. Overall, to what extent did you like this interaction technique?} \\
    \bottomrule
  \end{tabular}
\caption{Post-Questionnaire questions. The answers were in 7-point likert scale, and 
every question was asked on each tested method.}~\label{tab:2-3-4-postQ}
\end{table}

We replicated a typical home screen interface of VR headset (Figure \ref{fig:2-sshot}),
which consisted of 15 buttons of equal size.
Each task began by selecting the button in the middle (button no.7), and then 
one of the buttons except that middle button would turn red for selection.
Buttons turn green to indicate the success.
All Tasks began from the button no.7 to make the total travelled distance consistent for every participant. A session
consisted of a series of 14 tasks where users selected every button, except the middle button, 
in a random order. Each participant completed 3 sessions 
for each technique, the order of the techniques randomized based 
on a Latin square design \cite{latin}.
Post-questionnaire (Table \ref{tab:2-3-4-postQ}) and short interview followed after completing the experiment.
A total of 20 participants (16 males, all right-handed, 20's to 40's) were recruited, and 
three of them had participated in the previous experiments.

\subsection{Performance Result}
\begin{figure}[!t]
\centering
  \includegraphics[width=\columnwidth]{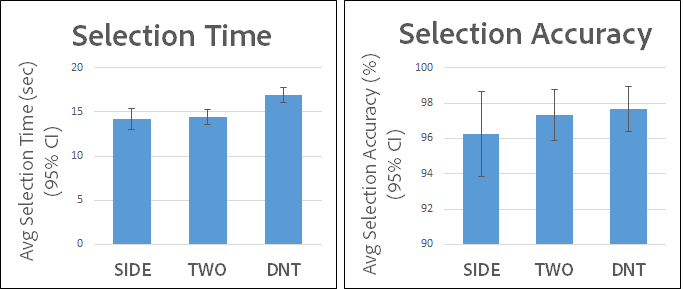}
  \caption{Average selection time and accuracy of each method: Side, Two-Fingers (TWO) and Drag-n-Tap (DNT). 
  Side and Two-Fingers were faster than Drag-n-Tap while Two-Fingers and Drag-n-Tap were more accurate than 
  Side. }~\label{fig:3-time-acc}
\end{figure}

We used repeated-measures one-way
ANOVA to examine the average selection time and accuracy, and a post-hoc
analysis using the paired t-tests with p-values adjusted by Holm-Bonferroni
correction \cite{Holm}.
Average selection time and accuracy are shown in Figure \ref{fig:3-time-acc}.
Users took 14.18, 14.42 and 16.87 seconds to
finish tasks using Side, Two-Fingers and Drag-n-Tap respectively 
with significant difference ($F_{2,38} = 14.312, p <
0.00005$). Pairwise t-tests revealed that Side ($t_{19} = -4.510, p < 0.001$) and Two-Fingers ($t_{19} = -4.386,
p < 0.001$) were faster than Drag-n-Tap, but no significant difference
between the first two ($t_{19} = -0.489, p = 0.631$). For accuracy,
Side achieved 96.25\%, Two-Fingers 97.32\%, and Drag-n-Tap 97.68\%, but no statistical
difference was found among the three ($F_{3,51} = 260.764, p = 0.502$).

\subsection{Insights from Users}
To analyze the likert scale data of the post-questionnaire,
we used Friedman rank sum test, and Wilcoxon signed rank test with
Holm-Bonferroni correction for post-hoc analysis.

\begin{figure}[!t]
\centering
  \includegraphics[width=\columnwidth]{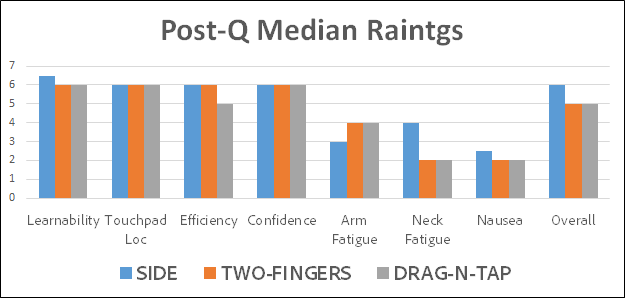}
  \caption{Medians of the post-questionnaire results for each method. Friedman test revealed
  significant differences in Learnability, Neck Fatigue and Nausea.}~\label{fig:3-likert-median}
\end{figure}

Figure \ref{fig:3-likert-median} shows the median scores of each post-questionnaire question.
We found significant difference in users' responses
for learnability ($F_{2} = 10.364, p < 0.01$), neck fatigue ($F_{2} = 13.636,
p < 0.005$) and nausea ($F_{2} = 18.2, p < 0.0005$).

\subsubsection{Learnability and Metaphors}
Side was reported to be easier to learn than Two-Fingers ($Z=2.840, p < 0.05$), 
but Drag-n-Tap showed no significant difference with the other two. This result 
corresponds with the interview feedback that Side was more natural 
and straightforward to use than the Two-Fingers (P2, P11, P12, P16).
Two-Fingers seems to be more difficult to learn as each finger serves 
a different role and users need to work them simultaneously. 
However, its resemblance to the motions of a mouse (P3, P13) or the trackpad on the laptop (P11) 
help them learn more quickly and once they get accustomed to,
Two-Fingers could be quicker than Side (P13). In a similar fashion, Drag-n-Tap 
also reminded them of a trackpad on laptop (P13) or a mouse (P17, P18, P19, P20), 
and this metaphor allowed the gesture be more intuitive to use.

\subsubsection{Neck Fatigue and Nausea}
Neck fatigue was raised as an issue with Side in relative to Two-Fingers
($Z=2.777, p < 0.05$) and Drag-n-Tap ($Z=2.777, p < 0.05$).
Since neck is less trained for delicate tasks than fingers, 
some experienced tension on their necks as they try to fixate them to target a small area (P2, P14).
Also, when selecting the buttons on 3x1 layout, some found it uncomfortable having to pan heads to that much of an angle (P1, P18). 
This might had effect on the nausea with Side as shown in the post-hoc analysis where 
Side was worse than Two-Fingers ($Z=2.699, p < 0.05$) and
Drag-n-Tap ($Z=2.848, p < 0.05$).

\subsubsection{Control over the Cursor}
Having a finger on the front touchpad throughout the experiment, users could know 
how the position of finger and cursor are mapped for Two-Fingers. This 
seemed to give users more confidence in making  selections (P15, P16) as 
they can have a full control over the cursor until the final tapping is signaled.
Stabilizing their necks than fingers seemed to be more difficult 
to achieve pinpoint precision.
Some even expressed their doubts in Side being able to handle delicate tasks, 
and that fingers are more trained for such tasks. 
For consecutive tasks, Two-Fingers seemed more suitable than Drag-n-Tap
since the actions of lifting the finger and re-tapping of Drag-n-Tap cause delays (P2, P4, P12, P13).
Two-Fingers, however, had more off-screen issues than Drag-n-Tap where 
users attempt to make a tap outside the touchpad. It is because they work with multiple fingers on a limited 
touchpad space for Two-Fingers. Users who experienced this off-screen issue 
with more frequently than others had tendency to prefer Drag-n-Tap (P5, P16). 

\section{Study 2: keyboard}
In this study, we explored how Side and Two-Fingers methods perform for VR
keyboard application. We used Two-Fingers instead of Drag-n-Tap because (1) its performance measures were better and 
(2) the hand gesture of Two-Fingers resembled that of the SWIFT keyboard which we 
want to explore in future.

\subsection{Study Design}
\begin{figure}[!t]
\centering
  \includegraphics[width=\columnwidth]{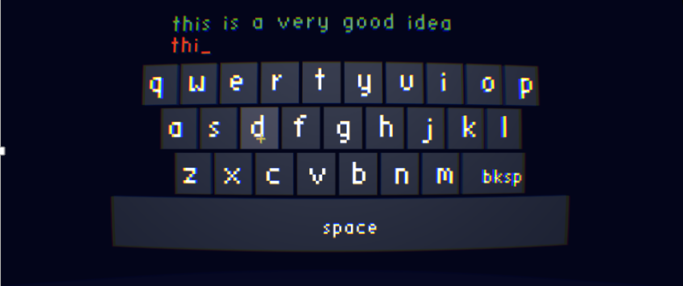}
  \caption{A snapshot of the VR keyboard used for the experiment. It illustrates a case where \textit{``this is a very good idea"} is the presented phrase (i.e. green text) and \textit{``thi"} is the current state of the transcription (i.e. orange text).}~\label{fig:4-sshot}
\end{figure}

Figure \ref{fig:4-sshot} shows a prototype VR keyboard we implemented for the experiment. It 
was in a QWERTY format, and we provided a sound feedback with a short clicking sound to help 
users to know whether a letter was selected.
We randomly chose 5 phrases (25 to 28 characters long) from MacKenzie and Soukoreff corpus \cite{MacPhrases} 
and had a participant to transcribe a given sentence as quickly and accurately as possible.
For a short practice session, each participant got to write two phrases, which were 
the same for all participants. A total of 25 people participated in the study (19 males, 20's to 40's, all right-handed, 
English skills all higher than 4 in a likert scale of 7 equal to a Native level).

\subsection{Performance Result}
For each phrase, we calculated the text-entry rate in words per minute (wpm) \cite{WPM2, WPM1}, and 
error rates based on Minimum String Distance \cite{ErrorRate2}.
\begin{align*}
\textit{Words-per-Minute} = \frac{|S - 1|}{M} 
\end{align*}
where $|T-1|$ is the length of the string to transcribe excluding the first character, 
and $S$ is the duration in seconds. 
The formula of calculating the error rate is as follows:
\begin{align*}
\textit{Error Rate} = \frac{MSD(A, B)}{\overline{S_A}} \times 100\%
\end{align*}
where $A$ and $B$ are the presented and transcribed text and $\overline{S_A}$ 
indicates the mean size of the alignments.
The average wpm of Side (11.738 wpm) was slightly higher than that of Two-Fingers (11.346
wpm), but with no significant difference ($t_{24} = 1.246, p = 0.225$).
The average error rate of Side ($0.24\%$) was lower than that of Two-Fingers ($0.57\%$), 
but again with no significant difference ($t_{24} = -1.913, p = 0.068$).

\subsubsection{Performance over Time}
\begin{figure}[!t]
\centering
  \includegraphics[width=\columnwidth]{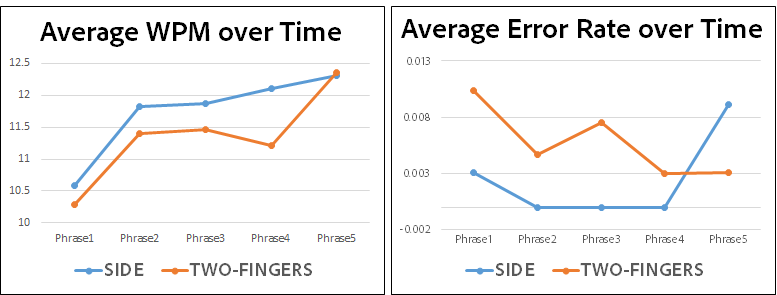}
  \caption{Average wpm and error rate measure over time for Side and Two-Fingers techniques. 
  For both wpm and error rate, Two-Fingers performed better at the fifth (last) phrase, hinting
  a sign of improvement of Two-Fingers as the time progresses.}~\label{fig:over-time-perf}
\end{figure}

We could discover potential benefits in using Two-Fingers 
in a longer term as we examined users' performance over time (Figure \ref{fig:over-time-perf}). We could observe increases in speed for both methods, and Two-Fingers outperformed Side at the fifth phrase. 
Similarly, the error rate of Side is bigger than Two-Fingers on the fifth.
The initial underperformance of Two-Fingers could have been due to its lower 
learnability relative to Side (P5) and/or the performance of Side could have become  
worse over time as the neck fatigue and nausea accumulate.
To confirm this hypothesis, we should conduct stress tests
and see how the two methods perform with longer sentences. 

\subsection{Insights from Users}
We conducted Wilcoxon signed rank test on
each post-questionnaire question.
Five out of the eight features were found to be
significantly different: efficiency, arm fatigue, neck fatigue, nausea and overall rating. 
The significant difference in overall rating statistically verified that users preferred Two-Fingers over Side ($Z = 2.541, p <
0.05$).

\begin{figure}[!t]
\centering
  \includegraphics[width=\columnwidth]{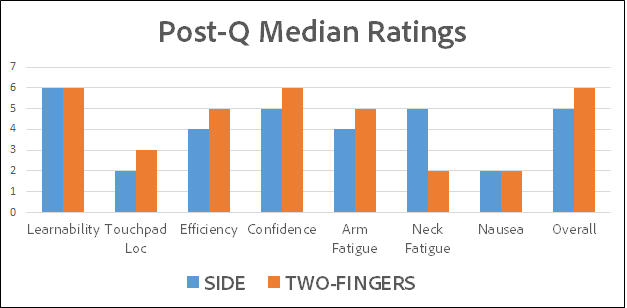}
  \caption{Medians of the post-questionnaire results for Side and Two-Fingers.
  Wilcoxon signed rank test revealed significant differences in Efficiency, Arm Fatigue, Neck Fatigue, 
  Nausea and Overall Rating.}~\label{fig:4-likert-median}\vspace{-4ex}
\end{figure}

\subsubsection{Efficiency of the Techniques to Text-entry}
From the likert scale analysis, we observed that users found Side technique
to be less efficient at text-input task ($Z = -2.013, p < 0.05$).
Five users answered that Side is inappropriate for the text-input task because 
moving head that much for a rudimentary task like text-input is tiresome, tedious and almost silly. 
Although moving head is generally faster,
fingers catch up quickly as you get used to the technique (P23).
Regarding the speed of interaction, people found moving head to 
be faster than moving their fingers across the touchpad to select a key on one end 
after selecting a key at the other end of keyboard (P11). This could be improved by accounting the velocity of finger movement to move the VR cursor, instead of simple linear mapping. Further discussion is done later in the paper.

\subsubsection{Fatigue and Nausea: Head vs. Hand}
As in the previous study, neck fatigue was a significant 
issue for Side ($Z = 3.642, p < 0.0005$). But with the keyboard, 
people also experienced arm fatigue for Two-Fingers ($Z = 3.642, p < 0.0005$).
Fifteen participants commented that they experienced arm/neck fatigue
in their interviews.
Users' preferences seemed to have been affected by this fatigue experience. For instance, five participants answered that they liked Two-Fingers because they regarded neck fatigue as a more severe issue while three liked Side better because they experienced arm fatigue with Two-Fingers. 
While users did not commented on getting neck fatigue for Two-Fingers, 
four users complaint on getting arm fatigue for Side as well since they have to hold their arms up anyway 
Also, since the head is in the course of constant movement, arm needs to stay 
aligned to the side touchpad not to lose its location, and this  
led to ergonomic annoyance. 
As the task gets more complex and courses of movements longer, the tolerance on 
moving hand/head becomes the major criteria on the preference between side and front.
In general, users preferred Side 
or had no preference 
for writing short sentences while preferred Two-Fingers for long sentences. 

\subsubsection{Precision}
Seven participants remarked that they were more prone to error with Side because  
they had to make a tap in a course of head movements. But if moving head and tapping did not synchronize well, they ended up selecting wrong keys: 
``when you are in the middle of writing sentences, the momentum (of head movement) builds up. And if the timing is not right, you tap on the side too quickly or too slowly and make a typo. (P24)"

Some participants replied it is harder to achieve high precision with Side than Two-Fingers with small and packed buttons. It was because it was harder to travel a small distance and make a pinpoint precision with a neck than with a hand. When they tried to make a pinpoint precision with Side, they often experienced muscle tension on their necks.
Also, they 
were more prone to losing the location of the side touchpad while using the keyboard as the head movements become more intense 
However, two users still remarked that gazing was a better means to achieve high precision (P12, P17). 

Two-Fingers, on the other hand, uses dexterity to control the cursor movement with ease and achieve higher precision (P6, P11, P23, P24, P25). Side may be faster, but when working with small buttons, clicking with higher precision becomes a priority than speed (P13).
Also, with Two-Fingers, they could hold their fingers down on the touchpad to have a full control over the cursor throughout the interaction. 
However, four participants commented that Two-Fingers also leads to error by accidentally tapping
twice during the interaction (P14, P15, P16, P21). 

\subsubsection{Limited Range of View}
Some users preferred the front touch because they 
could have an overall view of both the keyboard and the sentence they were transcribing.
With Side, however, the view became restricted as they move their heads to target at a specific key on the keyboard.
This inability to get an overall view of the keyboard and the sentences they are writing seemed to be a concern especially for longer sentences.

\subsubsection{Drag-n-Tap}
We asked users who also participated in the previous experiment how they would have liked Drag-n-Tap technique on a keyboard application.
A majority of them showed positive responses since the gesture 
is more straightforward (P4, P21, P22), less prone to the off-screen issue (P1, P13),
and perceived to support higher precision (P25).
But they also remarked that Drag-n-Tap will be less efficient 
in terms of performance than the other two techniques. 

\subsubsection{Like Neither}
In general, Two-Fingers was more preferred over Side with a statistical difference.
However, two participants responded that they had no strong preference over either of them and that neither of them looked suitable for text-entry for VR (P18, P19).
For VR text-entry, they said that they would rather use the voice recognition 
or a standard keyboard connected to the device if they can. They envisioned an interaction that can support as high fidelity as the standard keyboard or that on the phone to be developed for VR in future. 
But we also received a positive feedback that the front touch clearly has a room to be explored for further optimization, and that it can be improved to give more pleasing experience (P3). 

\section{Discussion}

\subsection{Which is more Efficient? (speed and precision)}
Quantitatively speaking, front touch performed better on speed than Side in Pe-Study 2, but 
there were no statistical differences between them on other study results.
We could also learn from the qualitative feedback that users could move faster with Side, but it is 
more difficult to control the cursor with high precision with Side than with Two-Fingers. It was because fixating to a small target area or moving a small amount of distance becomes more tedious with head than hand.

Even though it takes a little more time to learn for the front touch methods than Side, there is a higher chance of Two-Fingers outperforming Side once users become accustomed to (P9 from Study 1).
Also, since fingers are more trained at performing 
delicate tasks than necks, this makes the front touch interactions more suitable for UI's with higher complexity like the keyboard in a long term. Performance-over-time results from Study 2 (Figure \ref{fig:over-time-perf}) where Two-Fingers outperforms Side at the fifth phrase could be used to support this hypothesis. Further study should be conducted to find out the long-term performances of different techniques.

\subsection{Task Condition Dependent Preferences}
There is a general tendency to like the front touch interactions for complex 
tasks while gazing for simpler user interface. 
We examined the reasons behind 
these tendencies in a more specific manner, and summarized 
several factors that influenced the users' preference over different selection 
techniques. These factors are not independent of each other but rather influence one 
another. For instance, if the task becomes sequential, it is more likely to 
get tired, and the tolerance level for arm and neck might differ for 
short and long tasks. 

\subsubsection{Complexity of the User Interface}
Things like how small and spread out the buttons like in their field of view
largely influenced their choices. In general, people liked to use 
their hands more than heads as the buttons get smaller, and more spread. 
It is because users found it easier to make pinpoint accuracy 
with hands than head. Also, they would have 
to pan their heads in a wider angle if the buttons are more spread out, 
which can be offloaded if they can use their hand along with the head 
to reach for items out side of their field of view.

\subsubsection{Intensity and Length of Task}
For consecutive tasks like text-input, front touch interaction was preferred over Side users because it is easier to control the cursor with hand than with head. For Side, they were more prone to making errors by selecting keys too quickly or too slowly in the course of head movement. Also, some even felt that it was tedious and almost silly to move head that much in VR.

\subsubsection{Learnability of the techniques}
Although it was not indicated in any of the likert scale results, we observed users tend to take more time getting accustomed 
to Two-Fingers and Drag-n-Tap than Side. Unlike Side where they had to look at their target and tap, users took some time experimenting with the fingers postures for the front touch techniques.
Some quickly learned the posture of due to its resemblance 
of that of using a mouse, but some suffered at the beginning being 
accustomed to separating the roles of each finger and control the cursor.
Drag-n-Tap, in this sense, was found to be more intuitive than Two-Fingers.
A natural inclination to touch what you see (as shown from the Pre-Study 1 results) also helped users be accustomed to the new front touch interaction techniques.

\subsubsection{Engaging}
We received a number of commentaries from users that 
they found the gazing interaction to be entertaining and fun to use. 
Gazing and tapping movements seem to remind them of a game,
and made them to be more engaged in the interaction.
This unique feature of Side can make it an appropriate interaction 
for VR application games such as First-Person shoorter game where 
the goal is aiming at the target (gazing) and make a shot (tapping).

\subsubsection{Fatigue and Nausea}
As the complexity of user interface increases and tasks become longer, neck and arm fatigue issues 
start to play bigger roles.
The ergonomic position of the arm locked to the side of torso for Side technique 
helped users feel less arm fatigue than the floating gesture of front. 
But, the courses of head movements for consecutive tasks and the tense they experience 
on their muscles to achieve high precision on smaller targets was more severe on the neck. 
So, there is a trade-off made between neck and arm fatigue in choosing between the techniques. The preference depends on the level of tolerance on arm and neck fatigue, which differs from person to person.
Those who were more concerned about getting the neck tired would go for the front, while those more concerned 
about the arm for the side. 
But, Side is not completely free from arm fatigue issue either because users have to hold their arms up to touch the side pad anyhow.

Although we could summarize general tendencies based on the study results, these factors can vary at individual levels, 
and are difficult to measure them with absolute precision. Thus, it would be ideal to 
give a hybrid of these techniques, compensating for each other's limitations instead of 
one replacing another. Just as in \cite{grabandmani}, hybridized techniques would give distinct advantages in 
terms of use and efficiency.
Fortunately, these techniques can be supported together as one embedded interaction 
for a VR headset without having a technical clash. With multiple interaction options available, 
they can choose the right one depending on their use cases.

\subsection{Design Tips for the Front Touch Interface}
\subsubsection{Physicality of the Front Touchpad}
Smoother, lighter and more responsive surface would be needed for the 
front touch than our prototype device, to lessen the friction that 
builds up from moving across the touchpad for a long time. This 
especially could be an issue for Two-Fingers where one finger is in a constant contact with the surface of the pad. Thus, smoother, lighter and more responsive 
surface for the front pad to prevent such adherence should be supported. 

Furthermore, the prototype touch sensor device (i.e. Note4) was flat, but 
it will be worthwhile to experiment how the front interaction can be improved  
with the curved touchpad that covers the temples as demonstrated in the Figure 
\ref{fig:concept-art}. We hypothesize such design would prevent users from going outside the front touchpad.

\subsubsection{Coordinate Mapping between Touchpad and VR world}
For the experiments, the coordinates of front touchpad and 
that of virtual environment scene was linearly mapped. 
So, a consistent amount of distance was traveled in VR for every unit of finger movement 
on the touch surface. We can experiment with this design choice by 
applying the velocity of the finger movement to the VR. 
The faster the finger moves, the longer the distacnce the VR cursor travels. 
This feature should be designed carefully because there could be a trade-off made 
between the speed and accuracy. If users can make the cursor to travel a longer distance 
with a shorter finger movement, they could have more difficulty in achieving 
pinpoint precision. 

Furthermore, we can map the touchpad position onto VR world in two different ways: 
(1) absolute and (2) relative. In absolute mapping, the touchpad position 
is directly scaled to the VR world, while in relative mapping, the touchpad motion (changes in coordinates)
is added to the current VR world location.
We conducted a pilot study on ourselves to find which is more appropriate 
for different techniques. We noticed Two-Fingers caused more 
off-screen issues with relative mapping than absolute. It is because the initial touch 
position is mapped to the center of VR world. 
If users make initial touch slightly off from the center of 
the touchpad, they were more likely to encounter the off-screen issue. 
However, we believe this off-screen issue is the current test prototype's limitation. If the touch sensor covers
upper/lower and side (Figure \ref{fig:concept-art}), or simply larger area of the face, the off-screen issue may be negligible since users
first touch location is reasonably accurate (Pre-Study 2).

For absolute mapping, there is a jumping-cursor issue: a small fraction (below 10\%) of users reported 
confusion when the cursor suddenly jumped to the new touch position. 
These users appeared to be annoyed and perceive the whole interaction to be a ittle bit clumsy.
To improve this, we tested a hybrid of absolute and relative. We corrected the offset error by a fixed fraction each touch event.
The result was promising although we did not perform formal study.

We chose the relative mapping for Drag-n-Tap to avoid the jumping issue.
We chose the absolute positions for Two-Fingers to make the touch positions consistent for all users.

\subsubsection{Size of Buttons for Front Touch}
Users replied that they could have more control over the cursor with hand than head and move it to a precise location more flexibly. 
In this sense, we can infer that the front touch can afford the smaller sized buttons than the side technique.
Since the front touch can tolerate with small buttons, we can make the size of keys on keyboard smaller similar to the new netflix application in Figure \ref{fig:netflix}. This might even improve the text-input speed because users will be traveling less distance to move from one key to another.
These suggestions, however, will need to be empirically proven to find out the exact threshold value on the button size. 

\subsection{Application Areas}
The front touch interface opened a new possibilities of VR interactions, and its unique properties make it appropriate for a range of mass media applications 
that are being introduced to VR field.

With the introduction of the front touch and its application on text-entry tasks, 
we can now exchange messages with others more naturally, facilitating the social features of VR headsets. 
It also is appropriate for password input by shuffling the keys on the keyboard, and blocking others 
from decoding the input values.

Also, with the front touch interaction, World-fixed menu interface is now possible and 
users do not need to move their heads to direct the cursor anymore. Thus, they can 
now use the headsets at any comfortable posture, especially the postures when their 
heads are not free to move. For instance, users can now wear the headsets lying down 
and manipulate the menus to select a film to have personal theatre in their beds. 
With the interaction liberated from the head orientation, we can now afford a wider range of 
VR applications for various situations.

\subsection{Discussions on Intrinsic Counter Intuitivity}
Although Pre-Study 1 showed the front touch is intuitive, 
it is also counter intuitive in a sense that we are pushing a UI button 
from behind. To obtain an insight, we performed a short pilot study on ourselves. We tested two different ways of 
animating the button click: button is pushed (1) farther from eyes and (2) towards eyes. 
Note that (1) is opposite to the hand motion but consistent with conventional UI, while (2) is in the same direction as 
fingers but opposite to conventional UI. We chose (1) since we felt (1) is more natural.
In all user studies, no user pointed out this discrepancy in finger and button motions.
We obtained a strong sense that users appeared to be comfortable on (1). 

Unlike the traditional VR interaction where users ``reach out" their hands to 
interact with the objects, they instead ``push towards" their face for the front touch interaction. 
This less familar motion of the front touch initially confused some participants 
with prior VR experience, but they quickly adapted to it with their kinesthetic senses.
We also need to interact with items in multiple layers using the front touch interaction.
New gesture, such as pinching two fingers 
on the touchpad, can be developed to move the cursor across the z-index, or zoom-in and -out of the scene in VR. 
We can also hybridize the front touch with the traditional motion capture 
interaction, e.g. VR game user can use the gestural interaction during play, 
and the front touch to navigate through menus.

In spite of its limit, we believe we could open a new range of design possibilities by 
introducing the front touch interface to VR interaction. Its practical and simple characteristics make 
the front touch interface appropriate for the newly-emerging everyday application, such as movie-watching and 
social networking. 

\section{Penalty on the front touch}
When we were testing the side technique, we always took the touch sensor device (Note4)
off from the headset, which implies there was an extra weight penalty 
on the front touch techniques. Even though neck fatigue 
was found to be more severe with the side than the front, this extra weight 
could have influenced negatively on the performace of the front touch.
Also, there were occasional bluetooth disconnections, 
causing latency in sending the touch position data to VR system. 
We received feedbacks that these occasional bugs irritated them during the 
experiment, so there is a high chance of this technical issue negatively 
affecting on users' experience.

\section{Future Work}
We believe there are rooms for improvement to further develop the front touch interface. 
Among the four universal VR tasks - navigation, selection, manipulation and system control
\cite{FormalizeVE} - only the selection was explored in this paper.
Operations, such as drag-and-drop, pinching, and swiping, can be explored to study those other three mechanisms. 
Also, we hope to experiment the front touch interaction on View-fixed UI layout, 
to further find its unique advantages. We hypothesize that the results with the View-fixed UI layout 
will be better than that with the World-fixed, as noticed in the second pre-study.
We can also hybridize the front touch with other (e.g. side or gestural interaction) 
to find more enhanced interface.

For text-input application, we hope to develop the SwiftKey keyboard or 
design other soft keyboards optimized for VR environment.
It would be also interesting if we can find a natural way to 
use both hands instead of one with the front touch.

\section{Conclusion}
We introduced the front touch interface to virtual reality headsets, 
where users exploit their proprioceptions to interact with VR via a touchpad at 
the front of the headset. We explored its design space through a series of user studies, 
tasks including menu selection and keyboard application. 
Results demonstrate this new interaction to be intuitive as shown in the first Pre-Study, cause minimal nausea,
and was preferred by users when UI complexity increases as shown in the keyboard study.
Furthermore, we developed two front-touch 
interactions: Two-Fingers that supports quick selection, and Drag-n-Tap for accurate seleciton.
We esteem the front touch interface can be naturally embedded to 
the mobile VR headset, with the low-cost, low-weight and low power-budget characteristics of the 
touch screen. Follow-up studies are expected to further enhance the interface 
and shed light on a range of unexplored design possibilities.

\clearpage

\makeatletter
\g@addto@macro{\UrlBreaks}{\UrlOrds}
\makeatother
\bibliographystyle{SIGCHI-Reference-Format}
\bibliography{proceedings}

\end{document}